# A particle-in-cell simulation of rf breakdown

M Radmilović-Radjenović, H. S. Ko and J.K. Lee

*Electronics and Electrical Engineering Department, Pohang University of Science and Technology, Pohang, 790-784, S. Korea*

**Abstract.** Breakdown voltages of a capacitively coupled radio frequency argon discharge at 27 MHz are studied. We use a one-dimensional electrostatic PIC code to investigate the effect of changing the secondary emission properties of the electrodes on the breakdown voltage, particularly at low *pd* values. Simulation results are compared with the available experimental results and a satisfactory agreement is found.

## INTRODUCTION

It is well known that understanding of the non-equilibrium processes which occur in rf discharges during breakdown is of interest, both for industrial applications [1-4] and for a deeper understanding of fundamental plasma behavior [5-7]. In order to optimize plasma technological processes it is often necessary to know gas breakdown conditions in a discharge device. Therefore, it is of considerable interest to simulate and measure the breakdown curves in rf fields.

Typically Paschen curves are roughly "u" shaped with a minimum breakdown voltage at a specific *pd* and increasing voltages at both, increasing and decreasing values of *pd*. The breakdown voltage generally forms a fairly smooth curve, with the left hand branch of the curve being markedly steeper than the right hand branch. But, under certain circumstances inflection points and other changes in the slope of the breakdown curve have been measured [7]. It was found that the left-hand branch is a multivalued function of the gas pressure i.e., a single gas pressure corresponds to several breakdown voltages. The multivaluedness of the left-hand branch is seen both at small distances between the electrodes and at a large distances, while the right-hand branch has an inflection point, but only if the distance between the electrodes is small and the minimum on the curve lies at a pressure for which the electron-neutral collision rate is much larger than the frequency of the electric filed. The deviation of the left-hand branch of the curve into the high pressure region apparently occurs because of a decrease in the ionization cross-section. As the voltage is increased, the emission from the electrodes increases and the breakdown curve shifts into the low-pressure region.

## PARTICLE-IN-CELL SIMULATION

A one-dimensional electrostatic PIC code, with Monte Carlo collisions, to model a reactor with cylindrical electrodes is utilized. PIC modeling techniques have been well documented in previous publications [8,9] so only a brief description of the code is given here. The PIC method follows the transport of a number of superparticles. Each superparticle is composed of a large number of real particles-electrons and ions. The null collisions method [9] is used with isotropic scattering of the particles. Electrons can make ionization, excitation and elastic collisions and ions make charge exchange and elastic scattering collisions. Calculations were performed by using well established cross sections for argon [10]. Having in mind that the secondary emission processes are very important to determining the breakdown, in our simulations both electron impact and ion induced secondary emission processes are included.

## Electron impact secondary emission

In our simulations we use a Vaughan-based secondary electron production [11,12] that includes energy and angular dependence as well as a full emission spectrum including reflected and scattered primaries. This is a more accurate model of secondary electron production. The electron impact secondary emission may be represented by the secondary emission coefficient that is equal to the flux of the emitted electrons normalized to the initial flux. It is given by:

$$\delta(\varepsilon,\theta) = \delta_{max\ 0}(1+\frac{k_{s\delta}\theta^2}{2\pi})(we^{1-w})^k, \quad w=\frac{\varepsilon-\varepsilon_0}{\varepsilon_{max0}(1+k_{sw}\theta^2/2\pi)-\varepsilon_0} \quad (1)$$

where $\varepsilon$ is the incident energy of a particle and $\theta$ is its angle of incidence measured with respect to the surface normal, $\delta_{max}$ is the peak secondary emission coefficient corresponding to the energy $\varepsilon_{max}$ and normal incidence. The exponent $k$ is derived from a curve-fit analysis, $\varepsilon_0$ is the secondary emission threshold. $k_{s\delta}$ and $k_{sw}$ are a surface-smoothness parameters (both can vary between 0 for rough surfaces and 2 for polished surfaces).

## Ion induced secondary emission

Electrons released at the cathode travel the whole distance to the anode and produce more ionization than electrons created en rote. For this reason, the onset of breakdown is determined by $\gamma$–effects at the cathode. The secondary electron emission from a surface under the action of an ion is described by the coefficient quantifying the number of secondary electrons produced at the cathode per ion usually known as the electron yield per ion and denoted by $\gamma_i$. Although, this coefficient depends on the cathode material and the gas it was often assumed that $\gamma_i$ is constant [13-16].

In order to correct this deficiency first, we implement energy dependence of the coefficient $\gamma_i$ by using the expression that was suggested by Phelps and Petrovic [17]:

$$\gamma_i(\varepsilon_i) = \frac{0.006 \cdot \varepsilon_i}{1+(\varepsilon_i/10)} + 1.05 \cdot 10^{-4} \frac{(\varepsilon_i-80)^{1.2}}{1+(\varepsilon_i/8000)^{1.5}}, \quad (2)$$

where $\varepsilon_i$ is the incident energy of the ion.

Second, according to the angular dependence of the coefficient suggested by Thierberger et al. [18] and Thomas [19] we assume that the angular dependence of the electron yield per ion $\gamma_i$ is described by:

$$\gamma_i(\varepsilon_i, \theta) = \gamma_i(\varepsilon_i) \cos^{-1}\theta \qquad (3)$$

where $\theta$ is its angle of incidence measured with respect to the surface normal.

## RESULTS OF SIMULATIONS AND DISCUSSIONS

In order to determine the breakdown voltage, we use the fact that the breakdown is not an instantaneous phenomenon, it is occurs over a finite period of time which is determined by the balance between creation of charged species by ionization and their losses via collisional processes and diffusion to the walls. In figure 1 we show how breakdown voltage depends on the gas pressure. LAM experiment data for argon are presented by solid symbols and compared with our simulation results obtained using XPDC1 code and taking into account only energy dependence of the coefficient $\gamma_i$ (see eq. 2) and with simulation results obtained using XPDC1 code and taking into account both energy and angular dependence of the coefficient $\gamma_i$ (see eq. 3). Since $\cos\theta$ is less then or equal to 1 (eq. 3), simulation results obtained involving energy and angular dependence of the yield per ion are lower than simulation results obtained considering only energy dependence. As can be observed from figure 1, in both cases there are good agreements between the experimental and simulation results. We also present results obtained by XPDC1 code using the constant value for the electron yield per ion ($\gamma_i = 0.2$) as well as simulation results obtained using XPDP1 code also for the constant yield per ion ($\gamma_i = 0.2$). As can be observed form figure 1, simulation results obtained not taking into account energy and/or angular dependence of the electron yield per ion are in serious disagreements with the experimental results.

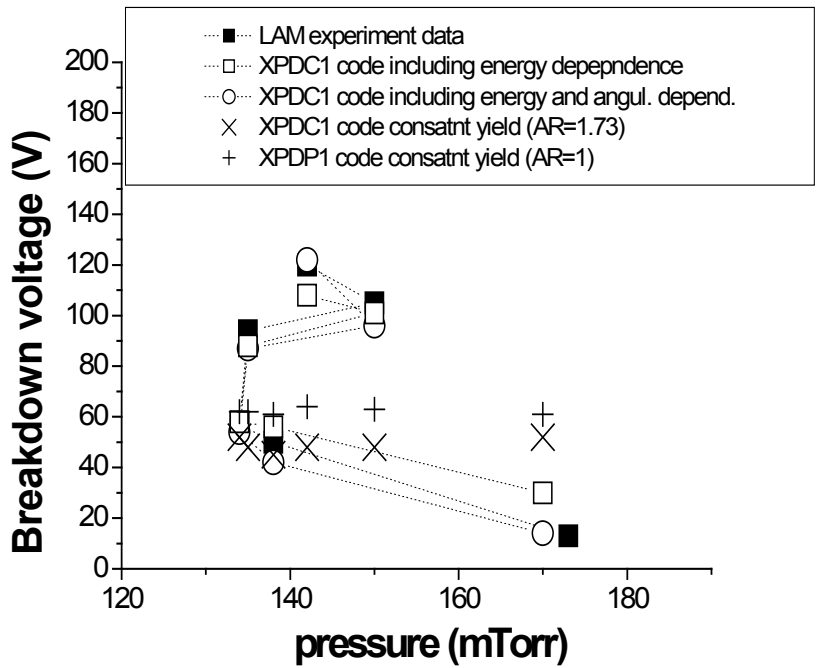

**Figure 1.** Breakdown voltage as a function of the pressure for argon.

## CONCLUSIONS

At high pressure and or large electrode separations, when the collision frequency is greater than the source frequency, breakdown conditions are dominated by volume processes and are relatively independent of surface conditions. At low pressure at small electrode separations, the loss rate of electrons to the walls is large so surface effects, in particular electron induced secondary emission, plays an important role in determining breakdown.

On the other hand, at high pressures where the electron oscillation amplitude in the axial direction in sufficiently small compared to the electrode separation, rf breakdown is mainly caused by the ionization within the gas. In the case of short electrode distances, the electron oscillation amplitude becomes comparable or larger than the electrode separation and subsequently the secondary electron emission may take place making ionization in the gas easier, lowering the breakdown voltage.

## REFERENCES

1. J.R. Ligenza, J.Appl.Phys. 36, 2703-2705 (1965).
2. A.C. Eckbreth and J.W. Davis, Appl.Phys.Lett. 21, 25-27 (1972).
3. D.L. Flamm et al., J.Vac.Sci.Technol. B1, 23-30 (1983).
4. K. Kobayashi et al., J.Appl.Phys. 59, 910-912 (1986).
5. T. Kihara, Rev.Mod.Phys. 24, 45-61 (1952).
6. V.A. Godyak and R.B. Piejak, Phys.Rev.Lett. 65(8), 996-999 (1990).
7. N.Yu. Kropotov, Sov.Tech.Phys.Lett. 14(2), 159-160 (1988).
8. C.K. Bridsall, IEEE Trans.Plasm.Sci. 19(2), 65-85 (1991).
9. V. Vahedi and M. Surendra, Comp.Phys.Commun. 87, 179-198 (1995).
10. J.K. Lee et al., IEEE Trans. On Plasma Science 32(1), 1-6 (2004).
11. J.R.M. Vaughan. IEEE Trans. 36, 1963-1967 (1989).
12. J.R.M. Vaughan, IEEE Trans. 40, 830 (1993).
13. J.P. Boeuf, Phys.Rev.A 36(6), 2782-2792 (1987).
14. T.H. Chung, H.J. Yoon, T.S. Kim and J.K. Lee, J. Phys. D:Appl. Phys. 29, 1014 (1996).
15. M. Soji and M. Sato, J.Phys.D:Appl.Phys. 32, 1640-1645 (1999).
16. H.B. Smith et al., Phys. of Plasm. 10(3), 875-881 (2003).
17. A.V. Phelps and Z.Lj. Petrović, Plasma Sources Sci. Technol. 8, R21-R44 (1999).
18. P. Thieberger et al., Phys. Review A 61, 1-11 (2000);
19. E.W. Thomas, Nuclear Fusion, Spec. Issue 94, 94 (1984).